%% file: main.tex
\documentclass[copyright,creativecommons]{eptcs}
\providecommand{\event}{SYNT 2017} 
\usepackage{amsthm}
\usepackage{amssymb}
\usepackage{mathtools}
\usepackage{temporal}
\usepackage{enumerate} 
\usepackage{color}
\usepackage{tikz}
\usepackage{paralist}
\usepackage{fixme,manfnt,mparhack}
\usepackage[T1]{fontenc}
\usepackage{enumerate}
\usepackage{xspace}
\usepackage{amssymb}
\usepackage{paralist}
\usepackage{wrapfig}
\usepackage{bbm}
\usepackage{cite}
\usepackage{booktabs}
\usepackage{array}
\usepackage{arydshln}

\usepackage{macros}

\providecommand{\thisvolume}[1]{this volume of {\sl Electronic
  Proceedings in Theoretical Computer Science}}

\definecolor{mygreen}{rgb}{0,0.6,0} 
\definecolor{mygray}{rgb}{0.9,0.9,0.9} 
\definecolor{mymauve}{rgb}{0.58,0,0.82}

\usepackage{listings}
\usepackage{xspace}

\input{preamble}

\title{The 4th Reactive Synthesis Competition (\syntcomp 2017): Benchmarks, Participants \& Results}
\author{Swen Jacobs
\institute{Saarland University\\Saarbr\"ucken, Germany}
\and
Nicolas Basset
\institute{Universit\'e Libre de Bruxelles\\ Brussels, Belgium}
\and
Roderick Bloem
\institute{Graz University of Technology \\ Graz, Austria}
\and 
Romain Brenguier
\institute{University of Oxford\\ Oxford, UK}
\and 
Maximilien Colange
\institute{LRDE, EPITA\\ Kremlin-Bic\^etre, France}
\and 
Peter Faymonville
\institute{Saarland University\\Saarbr\"ucken, Germany}
\and 
Bernd Finkbeiner
\institute{Saarland University\\Saarbr\"ucken, Germany}
\and
Ayrat Khalimov
\institute{Graz University of Technology \\ Graz, Austria}
\and 
Felix Klein
\institute{Saarland University\\Saarbr\"ucken, Germany}
\and
Thibaud Michaud
\institute{LRDE, EPITA\\ Kremlin-Bic\^etre, France}
\and
Guillermo A. P\'erez 
\institute{Universit\'e Libre de Bruxelles\\ Brussels, Belgium}
\and 
Jean-Fran\c{c}ois Raskin
\institute{Universit\'e Libre de Bruxelles\\ Brussels, Belgium}
\and
Ocan Sankur 
\institute{CNRS, Irisa\\Rennes, France}
\and 
Leander Tentrup
\institute{Saarland University\\Saarbr\"ucken, Germany}
}
\def\titlerunning{\syntcomp 2017}
\def\authorrunning{S. Jacobs et al.}
\begin{document}
\maketitle

\begin{abstract}
We report on the fourth reactive synthesis competition (\syntcomp 2017). We introduce two new benchmark classes that have been added to the SYNTCOMP library, and briefly describe the benchmark selection, evaluation scheme and the experimental setup of \syntcomp 2017. We present the participants of \syntcomp 2017, with a focus on changes with respect to the previous years and on the two completely new tools that have entered the competition. 
Finally, we present and analyze the results of our experimental evaluation, including a ranking of tools with respect to quantity and quality of solutions.
\end{abstract}

\input{intro}

\input{benchmarks}

\input{execution}

\input{participants}

\input{results}

\input{results-tlsf}

\input{conclusions}

\bibliographystyle{eptcs}
\bibliography{synthesis}
\end{document}

%% file: preamble.tex
\newcommand{\myparagraph}[1]{\smallskip \noindent {\em #1.}}

\newcommand{\available}[1]{is available at \url{#1}. Accessed August 2017.}

\newcommand{\demiurge}{\textsf{Demiurge}\xspace}

\newcommand{\syntcomp}{SYNTCOMP\xspace}

\newcommand{\abssynthe}{AbsSynthe\xspace}
\newcommand{\simpleBDD}{Simple BDD Solver\xspace}
\newcommand{\realizer}{Realizer\xspace}
\newcommand{\termitesat}{TermiteSAT\xspace}
\newcommand{\acaciaforaiger}{Acacia4Aiger\xspace}
\newcommand{\bowser}{BoWSer\xspace}
\newcommand{\ltlsynt}{\texttt{ltlsynt}\xspace}
\newcommand{\bosy}{BoSy\xspace}
\newcommand{\party}{\textsc{Party}\xspace}

\newcommand{\safetysynth}{SafetySynth\xspace}

\newcommand{\Abc}{\textsf{ABC}\xspace}

\newcommand{\subby}{\kern-0.2em\leftarrow\kern-0.2em}

%% file: intro.tex
\section{Introduction}
\label{sec:intro}
The reactive synthesis competition (\syntcomp) has been founded in 2014~\cite{SYNTCOMP14} with the goal to increase the impact of theoretical advancements in the automatic synthesis of reactive systems. Reactive synthesis is one of the major challenges of computer science since the definition of the problem more than 50 years ago~\cite{Church62}. A large body of theoretical results has been developed since then, but their impact on the practice of system design has been rather limited. \syntcomp is designed to foster research in scalable and user-friendly implementations of synthesis techniques by establishing a standard benchmark format, maintaining a challenging public benchmark library, and providing a \emph{dedicated and independent} platform for the comparison of tools under consistent experimental conditions.
	

Since its inception, \syntcomp is held annually, and is associated with the International Conference on Computer Aided
Verification (CAV) and the Workshop on Synthesis (SYNT), where the competition results are presented to the community.~\cite{SYNTCOMP14,SYNTCOMP15,SYNTCOMP16}
A design choice for the first two competitions was to focus on safety properties
  specified as monitor circuits in an extension of the AIGER format known from the hardware model checking competition~\cite{HWMCC14,SYNTCOMP-format}. \syntcomp 2016 introduced the first major extension of the competition by adding a new track that is based on properties in full linear temporal logic (LTL), given in the \emph{temporal logic synthesis format} (TLSF)~\cite{JacobsB16,JacobsK16}.
	
The organization team of \syntcomp 2017 consisted of R. Bloem and S. Jacobs.

\paragraph{Outline.} The rest of this paper describes the design, benchmarks, participants, and
results of \syntcomp 2017. In Section~\ref{sec:benchmarks}, we present two new benchmark classes that have been added to the \syntcomp library and give an overview of the benchmark set for \syntcomp 2017. In Section~\ref{sec:setup}, we describe the setup, rules and execution of the competition. In Section~\ref{sec:participants} we give an overview of the participants of \syntcomp 2017, focusing on changes compared to last year's participants. The experimental results are presented and analyzed in Section~\ref{sec:results}, before we end with some concluding remarks in Section~\ref{sec:conclusions}.

%% file: benchmarks.tex
\section{Benchmarks}
\label{sec:benchmarks}

In this section, we describe the benchmark library for \syntcomp 2017. We start by describing two new benchmark classes in TLSF, followed by a listing of the classes of benchmarks (in both TLSF and AIGER format) that have already been used in previous competitions. For more details on these existing benchmarks, we refer to the previous competition reports~\cite{SYNTCOMP14,SYNTCOMP15,SYNTCOMP16}.

\subsection{New Benchmark Set: Decomposed AMBA}
\label{sec:benchmarks-decomposed-amba}

This set of benchmarks has first been presented as an example in the presentation of TLSF~\cite{JacobsK16}. It describes the well-known AMBA bus controller, decomposed into eight components that can be synthesized independently. Out of these eight components, three are parameterized in the number of systems that access the shared bus, providing more challenging synthesis problems for larger parameter values. These benchmarks have been translated to TLSF by F. Klein.

\subsection{New Benchmark Set: Unrealizable Variants}
\label{sec:benchmarks-unreal-variants}

This set of benchmarks is based on a number of existing TLSF benchmark classes in the \syntcomp library: detector, full arbiter, load balancer, prioritized arbiter, round robin arbiter, simple arbiter (described below). All of the original benchmarks are parameterized in the number of systems that can send requests to the synthesized component. Moreover, all of them include a mutual exclusion property, and in this benchmark set have been modified in one of two ways to obtain unrealizable variants of the respective specification:
\begin{enumerate}
\item We add a requirement that the system has to serve multiple requests within a fixed number of steps, resulting in unrealizability due to a clash of mutual exclusion and the time limit on serving the requests. 
\item We add a requirement that forces the system to violate mutual exclusion not after a fixed time, but at some undetermined time in the future.
\end{enumerate}

\subsection{Existing Benchmarks: TLSF}
\label{sec:benchmarks-tlsf}

In addition to the new benchmarks described above, we briefly describe existing TLSF benchmarks. For more details consult the report on \syntcomp 2016~\cite{SYNTCOMP16} and the original sources. The existing benchmark library consists of the following classes of benchmarks:
\begin{itemize}
\item \textbf{Lily benchmark set:} the set of benchmarks originally included with the LTL synthesis tool \textsc{Lily}~\cite{JobstmannB06}. It includes $24$ benchmarks. 

\item \textbf{Acacia benchmark set:} the set of benchmarks originally included with the LTL synthesis tool Acacia+~\cite{bbfjr12}. It includes $65$ benchmarks.

\item \textbf{Parameterized detector:} specifies a component that raises its single output infinitely often if and only if all its inputs are raised infinitely often. Parameterized in the number of inputs.

\item \textbf{Parameterized arbiters:} four arbiter specifications of different complexity (simple arbiter, prioritized arbiter, round robin arbiter, full arbiter). Parameterized in the number of masters that the arbiter needs to serve.

\item \textbf{Parameterized AMBA bus controller:} essentially an arbiter with a large number of features, including prioritization and locking of the bus for a fixed or arbitrary number of steps~\cite{Jobstmann07b}. Parameterized in the number of masters that the controller has to serve.

\item \textbf{Parameterized load balancer:} a component that receives jobs and distributes them to a fixed number of servers~\cite{Ehlers12}. Parameterized in the number of servers that can handle the jobs.

\item \textbf{Parameterized generalized buffer:} a family of buffers that transmit data from a number of senders to two receivers, based on a handshake protocol and a FIFO queue that is used to store data~\cite{Jobstmann07b}. The benchmark is parameterized in the number of senders.

\item \textbf{Parameterized LTL to B\"uchi translation (LTL2DBA):} generation of deterministic B\"uchi automata that correspond to a specification taken from a set of parameterized LTL formulas~\cite{TianSDD15}.
\end{itemize}

\subsection{Existing Benchmarks: AIGER}
\label{sec:benchmarks-aiger}

We briefly describe the existing library of AIGER benchmarks. For more details, consult the previous competition reports~\cite{SYNTCOMP14,SYNTCOMP15,SYNTCOMP16} and the original sources. 

\begin{itemize}
\item \textbf{HWMCC benchmarks:} based on a subset of the benchmarks from HWMCC 2012 and HWMCC 2014~\cite{HWMCC14}, where a subset of the inputs have been declared as controllable, and a safe controller for these inputs should be synthesized.
This benchmark set contains $390$ benchmarks. 

\item \textbf{(Bounded) LTL to B\"uchi and LTL to parity translation (LTL2DBA/LTL2DPA):}
based on the benchmarks for LTL2DBA benchmarks from Section~\ref{sec:benchmarks-tlsf}, but including also the synthesis of parity automata, and additionally parameterized in the liveness-to-safety approximation. $62$ instances.

\item \textbf{Toy Examples:} a number of basic building blocks of circuits, such as an adder, a bitshifter, a counter, and a multiplier. The set consists of $176$ problem instances.

\item \textbf{AMBA:} a version of the bus controller specification for AMBA~\cite{Jobstmann07b}, parameterized in three dimensions (number of masters, type and precision of the liveness-to-safety approximation). The benchmark set contains $952$ instances.

\item \textbf{Genbuf:} a version of the generalized buffer specification~\cite{Jobstmann07b}, parameterized in the same way as the AMBA benchmarks. The set contains $866$ instances.

\item \textbf{LTL2AIG:} several sets of benchmarks that are based on the benchmark set of synthesis tool Acacia+~\cite{bbfjr12}, translated using the LTL2AIG tool~\cite{SYNTCOMP14}. This includes versions of the Lily, generalized buffer, and load balancer benchmarks mentioned above. $197$ problem instances.

\item \textbf{Factory Assembly Line:} a controller for two robot arms on an assembly line. This set contains $15$ problem instances.

\item \textbf{Moving Obstacle Evasion:} a moving robot that should evade a moving obstacle in two-dimensional space. The set consists of $16$ problem instances.

\item \textbf{Washing Cycle Scheduler:} a controller of a washing system, with water tanks that share pipes. Parameterized in the number of tanks, the maximum reaction delay, and the shared water pipes. The set contains $321$ instances.

\item \textbf{Driver Synthesis:} specifies a driver for a hard disk controller with respect to a given operating system model~\cite{RyzhykWKLRSV14}. Parameterized in the level of data abstraction, the precision of the liveness-to-safety approximation, and the simplification of the specification circuit by \Abc~\cite{abc}. $72$ instances.

\item \textbf{Huffman Encoder:} specifies a given Huffman decoder, for which a suitable encoder should be synthesized~\cite{Khalimov15}. Parameterized in the liveness-to-safety approximation, resulting in $5$ instances.  

\item \textbf{HyperLTL:} based on benchmark problems from HyperLTL model checking~\cite{FinkbeinerRS15}. The goal is to synthesize a witness for a given HyperLTL property.
This benchmark set contains $21$ instances.

\item \textbf{Matrix Multiplication:} asks for a circuit that performs a single matrix multiplication, or repeated multiplication with a subset of controllable inputs and an additional safety goal. Parameterized in the size of the input matrices, resulting in $354$ problem instances.
\end{itemize}

%% file: execution.tex
\section{Setup, Rules and Execution}
\label{sec:setup}

We give an overview of the setup, rules and execution of \syntcomp 2017. More details, and the reasoning behind different design choices, can be found in the first competition report~\cite{SYNTCOMP14} and previous work in which we also outlined plans for future extensions of the competition~\cite{JacobsB16}.

\subsection{General Rules}
\label{sec:rules}
Like in the previous year, there are two main tracks: one is based on safety specifications in AIGER format (in the following: AIGER/safety-track), and the other on full LTL specifications in TLSF (in the following: TLSF/LTL-track). The tracks are divided into subtracks for \emph{realizability checking }and \emph{synthesis}, and into two execution modes: \emph{sequential} (using a single core of the CPU) and \emph{parallel} (using up to $4$ cores). In the following, we start with rules that are common to both tracks, followed by rules that are specific to one of the tracks.

\paragraph{Submissions.}
Tools are submitted as source code, with instructions for installation and a description of the algorithms and optimizations used to solve the synthesis problem. Every tool can run in up to three configurations per subtrack and execution mode. After the initial submission, every tool is tested on a small set of benchmarks from the \syntcomp library, and authors are informed about any problems and can submit bugfixes.\footnote{Besides revealing bugs or shortcomings in the participating tools themselves, this year these tests have also revealed a number of problems in tools that are used as subprocedures. All of these problems have subsequently been addressed, thus providing an additional benefit of the competition to the broader community of formal methods research.}

\paragraph{Ranking Schemes.}
In all tracks, there is a ranking based on the number of correctly solved problems: a correct answer within the timeout of $3600$s is rewarded with one point for the solver, and a wrong answer is punished by subtracting $4$ points.
In the realizability tracks, correctness is determined by the realizability information stored in the files, if they have been used in previous competitions, or on a majority vote of the tools that solve the benchmark, otherwise. In the synthesis tracks, if the specification is realizable, then solution has to be model checked. This differs based on the input format, as explained below.

Furthermore, in synthesis tracks there is a ranking based on the \emph{quality} of the solution, measured by the number of gates in the produced AIGER circuit. To this end, the size $s$ of the solution is compared to the size $\mathit{ref}$ of a reference solution, which is either the smallest solution obtained by competition tools (during competition runs or special reference runs), or, if no previous solution exists, the smallest solution obtained by a tool in the current competition. The number of points obtained for a correct solution decreases logarithmically in the ratio of $s$ and $\mathit{ref}$, i.e., a correct answer (within the timebound) is rewarded with 

$$2 - log_{10}\left(\frac{s+1}{\mathit{ref}+1}\right)$$

points. Roughly, this means that for a solution with the same size as the best known solution, $2$ points are awarded. If the new solution is $10$ times bigger, $1$ point is awarded. If it is $10$ times smaller, $3$ points are awarded. A solution that is more than $100$ times bigger than $\mathit{ref}$ is awarded $0$ points. Note that since some synthesis problems can be solved without using a single gate, we cannot use the ratio $\frac{s}{\mathit{ref}}$, and use $\frac{s+1}{\mathit{ref}+1}$ instead.

\subsection{Specific Rules for AIGER/safety-Track}

\paragraph{Input Format.}
In the AIGER/safety-track, specifications are given in the Extended AIGER Format for Synthesis~\cite{SYNTCOMP-format,SYNTCOMP14}, modeling a single safety property.

\paragraph{Correctness of Solutions.}
In the synthesis subtrack, if the specification is realizable then tools must produce a solution in AIGER format. For \emph{syntactical correctness}, this solution must include the specification circuit, and must define how the inputs that are declared as \texttt{controllable} are computed from the \texttt{uncontrollable} inputs and the state variables of the circuit (for details, see the \syntcomp 2014 report~\cite{SYNTCOMP14}). 
To ensure also \emph{semantical correctness}, the solutions are additionally model checked, and only solutions that are both syntactically and semantically verified are counted as correct.
To facilitate verification, synthesis tools can optionally output an inductive invariant that witnesses the correctness of the solution, e.g., the winning region resulting from the attractor computation. Such an invariant is used \emph{in addition} to model checking, i.e., if the invariant check is inconclusive we fall back to full model checking.


\subsection{Specific Rules for TLSF/LTL-Track}

\paragraph{Input Format.}
In the TLSF/LTL-track, specifications are given in TLSF~\cite{JacobsK16}.
The organizers supply the \emph{synthesis format conversion} (SyFCo) tool\footnote{SyFCo is available at \url{https://github.com/reactive-systems/syfco}. Accessed August 2017.} that can be used to translate the specification to a large number of existing specification formats. Specifications are interpreted according to standard LTL semantics, with respect to realizability as a Mealy machine.

\paragraph{Correctness of Solutions.}
In the synthesis subtrack, tools have to produce a solution in AIGER format if the specification is realizable. For \emph{syntactical correctness}, the sets of inputs and outputs of the specification must be identical to the sets of inputs and outputs of the solution. To verify \emph{semantical correctness}, the solutions are additionally model checked against the specification with existing model checking tools. Only a solution that can be verified both syntactically and semantically is counted as correct.


\subsection{Selection of Benchmarks}
\label{sec:selection}

\paragraph{AIGER/safety-track.} In the AIGER-based track, the selection of benchmarks is based on information about the realizability and difficulty of benchmark problems that has been obtained from the results of previous competitions. This information is stored inside the benchmark files, as described in the \syntcomp 2015 report~\cite{SYNTCOMP15}.
For realizable specifications, we additionally determined the smallest known solution, stored as a \emph{reference size}. 

In \syntcomp 2017, we used the same benchmark classes and the same number of problems per class as in the previous year, but we randomly exchanged some of the problems within any given class (for classes that contain more problems than are selected for the competition), while preserving an even distribution of difficulty over the given class.
The number of selected problems from each category (cp. Section~\ref{sec:benchmarks-aiger}) is given in Table~\ref{tab:selected-benchmarks}. 

\begin{table}[h]
\caption{Number of selected Benchmarks per Category, AIGER/safety-track}
\label{tab:selected-benchmarks}
\centering
\def\arraystretch{1.2}
\begin{tabular}{ll|ll}
Category & Benchmarks & Category & Benchmarks\\
\hline
AMBA & 16 & Genbuf (LTL2AIG) & 8\\
(Washing) Cycle Scheduling & 16 & Add (Toy Examples)& 5\\
Demo (LTL2AIG)& 16 & Count (Toy Examples)& 5\\
Driver Synthesis & 16 & Bitshift (Toy Examples)& 5\\
Factory Assembly Line & 15 & Mult (Toy Examples)& 5\\
Genbuf & 16 & Mv/Mvs (Toy Examples)& 5\\
HWMCC & 16 & Stay (Toy Examples)& 5\\
HyperLTL & 16 & Huffman Encoder & 5\\
Load Balancer (LTL2AIG)& 16\\
LTL2DBA/LTL2DPA & 16\\
Moving Obstacle & 16 & \\
Matrix Multiplication & 16 & {\bf Total:} & {\bf 234}\\
\end{tabular}
\end{table}

\paragraph{TLSF/LTL-track.} In the TLSF-based track, for realizability checking we used all $24$ of the non-parameterized benchmarks from the Lily benchmark set, and $64$ from the Acacia benchmark set. Additionally, we used $6$ instances of each of the parameterized benchmarks. Overall, this amounts to $244$ problem instances.

\subsection{Execution}
\label{sec:execution}
Like in the last two years, \syntcomp 2017 was run at Saarland University, on a set of identical machines with a single quad-core Intel Xeon processor (E3-1271 v3, 3.6GHz) and 32 GB RAM (PC1600, ECC), running a GNU/Linux system. Each node has a local 480 GB SSD that can be used as temporary storage.

Also like in previous years, the competition was organized on the EDACC 
platform~\cite{BalintDGGKR11}, with a very similar setup.
To ensure a high comparability and reproducability of our results, a complete machine
was reserved for each job, i.e., one synthesis tool (configuration) running 
one benchmark. Olivier Roussel's 
\texttt{runsolver}~\cite{Roussel11}
was used to run each job and to measure CPU time and wall time, as well as 
enforcing timeouts. As all nodes are 
identical and no other tasks were run in parallel, no other limits than a 
timeout of $3600$ seconds (CPU time in sequential mode, wall time in
parallel mode) per benchmark was set.
Like last year, we used wrapper scripts to execute solvers that did not conform completely with the output format specified by the competition, e.g., to filter extra information that was displayed in addition to the specified output.

The model checker used for checking correctness of solutions for the AIGER/Safety track is IIMC\footnote{IIMC is available at \url{ftp://vlsi.colorado.edu/pub/iimc/}. Accessed August 2017.} in version 2.0. For solvers that supply an inductive invariant as a witness of correctness, we used a BDD-based invariant check to check correctness\footnote{Available from the \syntcomp repository at \url{https://bitbucket.org/swenjacobs/syntcomp/src}, in subdirectory \texttt{tools/WinRegCheck}. Accessed August 2017.}, and used full model-checking as a fallback solution if the invariant check failed.

For the TLSF/LTL track, the model checker used was V3\footnote{V3 is available at \url{https://github.com/chengyinwu/V3}. Accessed August 2017.}~\cite{WuWLH14}.

%% file: participants.tex
\section{Participants}
\label{sec:participants}
Overall, ten tools were entered into \syntcomp 2017: five in the AIGER/safety-track, and five in the TLSF/LTL-track. We briefly describe the participants and give pointers to additional information.

\subsection{AIGER/safety-Track}

This track had five participants in 2017, which we briefly describe in the following. 
For additional details on the implemented techniques and optimizations, we refer to the previous \syntcomp reports~\cite{SYNTCOMP14,SYNTCOMP15,SYNTCOMP16}. 

\subsubsection{Updated Tool: Swiss \abssynthe v2.1} 
\abssynthe was submitted by R. Brenguier, G. A. P\'erez, J.-F. Raskin, and O. Sankur, and competed in both the realizability and the synthesis track. It implements the classical backward-fixpoint-based approach to solving safety games using BDDs. As additional features, it supports decomposition of the problem into independent sub-games, as well as an abstraction approach~\cite{BrenguierPRS14,BrenguierPRS15}. This year, \abssynthe contains a new approach to compositionality, where the problem is not separated into as many sub-games as possible, but rather merges some of the smaller sub-games in the hope that this will yield more useful information about the overall problem.
It competes in the following sequential (SCx) and parallel (PCx) configurations:
\begin{itemize}
\item (SC1) uses a standard BDD-based fixpoint computation with several optimizations, but without compositionality or abstraction,
\item (SC2) uses an abstraction algorithm, but no compositionality, and 
\item (SC3) uses a compositional algorithm, combined with an abstraction method. This is the only sequential configuration that changed this year, incorporating the new approach to compositionality mentioned above.
\item (PC1) runs the three sequential configurations in parallel, plus one additional configuration that uses abstraction with fixed threshold, but no compositionality,
\item (PC2) runs four copies of (PC1), only modified in the BDD reordering technique, and
\item (PC3) runs four copies of (PC2), with the same set of different reordering techniques.
\end{itemize}

\paragraph{Implementation, Availability}
The source code of \abssynthe is is available at
\url{https://github.com/gaperez64/AbsSynthe/tree/native-dev-par}.

\subsubsection{Re-entered: \demiurge 1.2.0} 
\demiurge was submitted by R. K\"onighofer 
and M. Seidl,
and competed in both the realizability and the synthesis track.
\demiurge implements different SAT-based methods for solving the reactive synthesis problem~\cite{BloemKS14,SeidlK14}. In the competition, three of these methods are used --- one of them as the only method in sequential configuration (D1real) , and a combination of all three methods in parallel configuration (P3real). This year, \demiurge competed in the same version as last two years. 

\paragraph{Implementation, Availability}
{ \sloppy The source code of \demiurge is
available at {\url{%
https://www.iaik.tugraz.at/content/research/opensource/demiurge/}}
under the GNU LGPL version 3. }

\subsubsection{Re-entered: \safetysynth}
\safetysynth was submitted by L. Tentrup, 
and competed in both the realizability and the synthesis track. \safetysynth is a re-implementation of \realizer that implements the standard BDD-based algorithm for safety games, using the optimizations that were most beneficial for BDD-based tools in \syntcomp 2014 and 2015~\cite{SYNTCOMP14,SYNTCOMP15}. It competed in the same version as in the previous year, with configurations (basic) and (alternative) that only differ in the BDD reordering heuristic. 

\paragraph{Implementation, Availability}
The source code of \safetysynth is available online at: \url{https://www.react.uni-saarland.de/tools/safetysynth/}. 

\subsubsection{Re-entered: \simpleBDD}
\simpleBDD was submitted by L. Ryzhyk and A. Walker, and competed in the realizability track.
\simpleBDD implements the standard BDD-based algorithm for safety games, including a large number of optimizations in configuration (basic). The other two configurations additionally implement an abstraction-refinement approach inspired by de Alfaro and Roy~\cite{dealfaro} in two variants: with overapproximation of the winning region in configuration (abs1), or with both over- and underapproximation in (abs2). The version entered into \syntcomp 2017 is the same as last year.

\paragraph{Implementation, Availability}
The source code of \simpleBDD is available online at
\url{https://github.com/adamwalker/syntcomp}.

\subsubsection{Re-entered: \termitesat}
\termitesat was submitted by A. Legg, N. Narodytska and L. Ryzhyk, and competed in the realizability track. \termitesat implements a novel SAT-based method for synthesis of safety specifications based on Craig interpolation. The only configuration in sequential mode implements this new approach, and the parallel configurations (portfolio) and (hybrid) run the new algorithm alongside one of the algorithms of \simpleBDD~\cite{LeggNR16}, where in (hybrid) there is even communication of intermediate results between the different algorithms.

\paragraph{Implementation, Availability}
The source code of \termitesat is available online at: \url{https://github.com/alexlegg/TermiteSAT}.

\subsection{TLSF/LTL-Track}
\label{sec:participants-TLSF}

This track had five participants in 2017, two of which have not participated in previous competitions. All tools competed in both the realizability and the synthesis track. We describe the implemented techniques and optimizations of the new tools, followed by a brief overview of the updated tools. For additional details on the latter, we refer to the report of \syntcomp 2016~\cite{SYNTCOMP16}. 

\subsubsection{New Entrant: \bowser}
\bowser was submitted by B. Finkbeiner and F. Klein. It implements different extensions of the bounded synthesis approach~\cite{Finkbeiner13} 
that solves the LTL synthesis problem by first translating the specification into a universal co-B\"uchi automaton, and then encoding acceptance of a transition system with bounded number of states into a constraint system. In this case, the constraints are encoded into propositional logic, i.e., a SAT problem. A solution to this SAT problem, i.e., a satisfying assignment, then represents a transition system that satisfies the original specification. To also check for unrealizability of a formula, the dual problem of whether there exists a winning strategy for the environment is also encoded into SAT. 

For the synthesis of solutions in the basic configuration the satisfying assignment from the SAT solver is encoded into an AIGER circuit, and then handed to Yosis for simplification.
As a first extension, \bowser implements \emph{bounded cycle synthesis}~\cite{FinkbeinerK16}, which restricts the structure of the solution with respect to the number of cycles in the transition system. To this end, it additionally encodes into SAT the existence of a witness structure that guarantees that the number of cycles in the implementation is smaller than a given bound (according to the approach of Tiernan~\cite{Tiernan70}).

In addition, \bowser supports another encoding into SAT that targets AIGER circuits more directly, where the numbers of gates and latches can be bounded independently.

\bowser competed in the following configurations:
\begin{itemize}
\item configuration (c0) implements bounded synthesis in the basic version mentioned above,
\item configuration (c1) implements bounded cycle synthesis on top of bounded synthesis, i.e., in a first step it searches for a solution with a bounded number of states, and if that exists, it additionally bounds the number of cycles
\item configuration (c2) also implements bounded cycle synthesis on top of bounded synthesis, with the additional direct encoding into bounded AIGER circuits mentioned above.
\end{itemize}

In sequential mode, these configurations spawn multiple threads that are executed on a single CPU
core. The parallel configurations are essentially identical, except that
the threads are distributed to multiple cores.

\paragraph{Implementation, Availability}
\bowser is implemented in Haskell, and uses LTL3BA\footnote{LTL3BA \available{https://sourceforge.net/projects/ltl3ba/}}~\cite{ltl3ba} to convert specifications into automata, and MapleSAT\footnote{MapleSAT \available{https://sites.google.com/a/gsd.uwaterloo.ca/maplesat/}}~\cite{LiangGPC16} to solve SAT queries. For circuit generation, it uses the Yosis framework\footnote{Yosis \available{http://www.clifford.at/yosys/}}~\cite{Glaser2014}.
The website of \bowser is \url{https://www.react.uni-saarland.de/tools/bowser/}, where the source code will be made available soon.

\subsubsection{New Entrant: \ltlsynt}
\ltlsynt was submitted by M. Colange and T. Michaud and competed in a single configuration in both the sequential realizability and sequential synthesis tracks.
 
To solve the synthesis problem, \ltlsynt uses a translation to parity games. As a first step, the input LTL formula is translated into an $\omega$-automaton with a transition-based generalized B\"uchi acceptance condition. The resulting automata are more concise than classical state-based B\"uchi automata, which is important to make subsequent steps more efficient. Then, the automaton is simplified according to several heuristics, for example by removing non-accepting strongly connected components or redundant acceptance marks.~\cite{Duret16}
To separate the actions of the environment and the controller, each transition of the obtained automaton is split in two consecutive transitions, corresponding to the uncontrollable inputs and the controllable inputs of the original transition, respectively. 
The (non-deterministic) split automaton is then translated into a deterministic parity automaton, which can be interpreted as a turn-based parity game that is equivalent to the original synthesis problem. Determinism is key in preserving the semantics of the synthesis problem: every action of the environment can be answered by the controller so that the resulting run satisfies the LTL specification. Here, the controller wins the parity game (recall that such games are determined~\cite{Martin75}) if and only if the original instance of the reactive synthesis problem has a solution. 

\ltlsynt implements two algorithms that solve such parity games: the well-known recursive algorithm by Zielonka~\cite{Zielonka98}, and the recent quasi-polynomial time algorithm by Calude et al.~\cite{CaludeJKL017}. Note that the parity automata (and hence the parity games) produced by Spot are transition-based: priorities label transitions, not states. Again, this allows more concise automata, but required to adapt both algorithms to fit this class of automata. The default algorithm of \ltlsynt is Zielonka's, since in preliminary experiments it outperformed the algorithm of Calude on the benchmarks of SYNTCOMP 2016~\cite{SYNTCOMP16}. In fact, the experiments also showed that the bottleneck of the procedure is the determinization step, and not the resolution of the parity game.

A winning strategy for the controller in the parity game defines a satisfying implementation of the controller in the synthesis problem.
Since parity games are memoryless, or more precisely positional, a winning strategy can be obtained by removing edges of the parity game, so that each controlled state has exactly one outgoing edge. After reversing the splitting operation by merging consecutive transitions in this strategy, it can be straightforwardly encoded into an AIGER circuit. Binary Decision Diagrams (BDDs) are used to represent sets of atomic proposition, allowing some simplifications in the expression of outputs and latches. \ltlsynt also uses BDDs
to cache the expressions represented by AIGER variables to avoid adding redundant gates. In contrast to its competitors, \ltlsynt does not use external tools such as \Abc or Yosis for the encoding of solutions into AIGER.

\paragraph{Implementation, Availability}
\ltlsynt is implemented in C++ and integrated into a branch of the Spot automata library~\cite{Duret16}, which is used for translation of the specification into automata, and for manipulation of automata. Spot also integrates the BDD library BuDDy and the SAT solver PicoSAT. 
The source code of \ltlsynt is available in branch \texttt{tm/ltlsynt-pg} of the GIT repository of Spot at \url{https://gitlab.lrde.epita.fr/spot/spot.git}.

\subsubsection{Updated Tool: \acaciaforaiger}
\acaciaforaiger was submitted by R. Brenguier, G. A. P\'erez, J.-F. Raskin, and O. Sankur. It is an extension of the reactive synthesis tool Acacia+\footnote{Acacia+ \available{http://lit2.ulb.ac.be/acaciaplus/}}, which solves the reactive synthesis problem for LTL specifications by a reduction to safety games, which are then solved efficiently by symbolic algorithms based on an \emph{antichain} representation of the sets of states~\cite{bbfjr12}.
For \syntcomp 2017, Acacia+ has been extended with a parallel mode that searches independently for a system implementation and a counter-strategy for the environment. 

\paragraph{Implementation, Availability}
\acaciaforaiger is implemented in Python and C. It uses the AaPAL library\footnote{AaPAL \available{http://lit2.ulb.ac.be/aapal/}} for the manipulation of antichains, and the Speculoos tools~\footnote{Speculoos \available{https://github.com/romainbrenguier/Speculoos}} to generate AIGER circuits. The source code of \acaciaforaiger is available online at: \url{https://github.com/gaperez64/acacia4aiger}.

\subsubsection{Updated Tool: \bosy} 
\bosy was submitted by P. Faymonville, B. Finkbeiner and L. Tentrup, and competed in both the realizability and the synthesis track. \bosy uses the \emph{bounded synthesis} approach~\cite{Finkbeiner13} with an encoding into quantified Boolean formulas (QBF), as described by Faymonville et al.~\cite{FaymonvilleFRT17,FaymonvilleFT17}. 
To detect unrealizability, the existence of a bounded strategy of the environment to falsify the specification is encoded in a similar way, and checked in parallel. If no solution is found for a given bound on the size of the implementation, then the bound is increased in an exponential way.

The resulting QBF formulas are first simplified, using the QBF preprocessor bloqqer\footnote{Bloqqer \available{http://fmv.jku.at/bloqqer}}. In realizability mode, the simplified formula is then  directly solved, using the QBF solver RAReQS~\cite{JanotaKMC16}. In synthesis mode, \bosy uses a combination of RAReQS the certifying QBF solver QuAbS~\cite{Tentrup16}, and the certificate returned by QuAbS represents a solution to the synthesis problem. This solution is then converted into AIGER format, and further simplified using the \Abc framework. 

Two configurations of \bosy competed in \syntcomp 2017, differing in the translation from LTL to automata: configuration (spot) uses the Spot framework~\cite{Duret16}, configuration (ltl3ba) uses LTL3BA~\cite{ltl3ba} for this task. Both configurations support a parallel mode, if more than one core is available. \bosy supports both Mealy and Moore semantics natively, as well as the extraction of a winning strategy for the environment in case of unrealizable specifications.

\paragraph{Implementation, Availability.}
\bosy is written in Swift. It uses LTL3BA or Spot\footnote{Spot \available{https://spot.lrde.epita.fr}} to convert LTL specifications into B\"uchi automata. It uses bloqqer, RAReQS\footnote{RAReQS \available{http://sat.inesc-id.pt/~mikolas/sw/areqs/}} and QuAbs\footnote{QuAbs \available{https://www.react.uni-saarland.de/tools/quabs/}} to solve QBF constraints, and \Abc to simplify solutions.

The code is available online at: \url{https://www.react.uni-saarland.de/tools/bosy/}.

\subsubsection{Updated Tool: \party}

\party was submitted by A. Khalimov, and competed in both the realizability and the synthesis track, with three configurations (int, bool, aiger) in sequential mode and one configuration (portfolio) in parallel mode. The tool and the basic algorithms and optimizations it implements have been described in more detail by Khalimov et al.~\cite{KhalimovJB13,KhalimovJB13a}.

\party uses the bounded synthesis approach~\cite{Finkbeiner13} for solving the LTL synthesis problem.
To detect unrealizability, it uses the standard approach of synthesizing the dualized specification, where the synthesizer searches for a strategy for the environment to falsify the specification.
On a given benchmark, PARTY starts two threads to check realizability and unrealizability.
The check for unrealizability is limited to $1200$ seconds, giving more resources to the realizability check.

\party competed in four configurations:

\begin{itemize}
\item \party (int) is a re-entry from the previous year, with minor updates.
In this configuration, the synthesis problem is encoded into SMT satisfiability as follows.
The LTL formula is first translated into a universal co-B\"uchi automaton (UCA) using Spot~\cite{Duret16}.
Then the emptiness of the product of the automaton and an unknown fixed-size system is encoded into SMT satisfiability.
If the SMT query (in logic QF\_UFLRA) is satisfiable,
then \party (int) extracts the state machine,
encodes it into Verilog, and translates the Verilog file into AIGER using Yosis.
If the SMT query is unsatisfiable, it increases the system size and repeats the previous steps.

\item In \party (bool) a given LTL formula is translated into a UCA, similar to the approach of \party (int).
Then, in contrast, \party (bool) translates the UCA into a \emph{$k$-liveness automaton},
where the number of visits to any final state of the original UCA is limited by a heuristically chosen bound.
I.e., such an automaton approximates liveness properties up to some bound.
The rest is as before, we only note that the resulting SMT query is in the simpler logic QF\_UF.

\item \party (aiger) is a new entrant this year.
Similarly to the two previous configurations,
it follows the bounded synthesis approach~\cite{Finkbeiner13},
but the games-based one, not the SMT-based one.
First, a given LTL formula is reduced into a $k$-liveness automaton as before.
Since the result is a safety automaton, it can be determinized,
and translated into a safety game that is then encoded in Verilog.
(We use the basic subset construction for determinization,
 where for each automaton state we introduce one memory latch.)
Such a Verilog circuit is then converted into AIGER using Yosis
(this translation takes \emph{a lot} of time,
 likely due to optimizations that Yosis applies).
The resulting AIGER synthesis problem is solved with the SDF solver
that participated in SYNTCOMP 2016~\cite{SYNTCOMP16}.
SDF was slightly modified to produce AIGER circuits adhering to the TLSF track requirements.
If SDF does not find the solution, we increase the parameter $k$ and repeat.

\item
\party (portfolio) runs $5$ solvers in parallel:
(1) \party (aiger),
(2) \party (aiger) with formula strengthening,
(3) \party (aiger) on dualized specifications,
(4) \party (int) with system sizes from $1$ to $8$ (motivation: specifications requiring more than $8$ states are unsolvable anyway), and
(5) \party (bool) with fixed system size of $16$ (motivation: solving many unsatisfiable queries for small system sizes takes a significant time).
\party (portfolio) reports the first solution that any of the solvers find. The choice of these five solvers was based on the known strengths and weaknesses of the individual algorithms, and not on an analysis of their performance (individually or as portfolio) on the SYNTCOMP benchmark set.
\end{itemize}

\paragraph{Implementation, Availability.}
\party is written in Python.
It uses Spot to convert LTL specifications into B\"uchi automata,
Z3\footnote{Z3 \available{https://github.com/Z3Prover/z3}}~\cite{Moura08} to solve SMT constraints, and Yosis to translate Verilog into AIGER.
The code is available online at: \url{https://github.com/5nizza/party-elli}, branch syntcomp17.

%% file: results.tex
\section{Experimental Results}
\label{sec:results}

We present the results of \syntcomp 2017, separated into the AIGER/safety-track and the TLSF/LTL-track. Both main tracks are separated into realizability and synthesis subtracks, and parallel and sequential execution modes.
Detailed results of the competition are also directly accessible via the web-frontend of our instance of the EDACC platform at \url{http://syntcomp.cs.uni-saarland.de}.

\subsection{AIGER/safety-Track: Realizability}

In the track for realizability checking of safety specifications in AIGER format, $5$ tools competed on $234$ benchmark instances, selected from the different benchmark categories as explained in Section~\ref{sec:selection}. Overall, $16$ different configurations entered this track, with $10$ using sequential execution mode and $6$ using parallel mode. In the following, we compare the results of these $16$ configurations on the $234$ benchmarks selected for \syntcomp 2017.

We first restrict the evaluation of results to purely sequential tools, then extend it to include also the parallel versions, and finally give a brief analysis of the results.

\paragraph{Sequential Mode.}
In sequential mode, \abssynthe competed with three configurations (seq1, seq2 and seq3), \demiurge with one configuration (D1real), \simpleBDD with three configurations (basic, abs1, abs2), \safetysynth with two configurations (basic and alternative), as well as \termitesat with one configuration.

The number of solved instances per configuration, as well as the number of uniquely solved instances, are given in Table~\ref{tab:results-realseq}. No tool could solve more than $171$ out of the $234$ instances, or about $73\%$ of the benchmark set. $22$ instances could not be solved by any tool within the timeout. 

\begin{table}[h]
\caption{Results: AIGER Realizability (sequential mode only)}
\label{tab:results-realseq}
\centering
\def\arraystretch{1.3}
{\sffamily \small
\begin{tabular}{@{}llll@{}}
\toprule
Tool & (configuration)		& Solved  &  Unique \\ 
\midrule
\simpleBDD & (abs1) & 171 		& 0 \\
\safetysynth & (basic)      & 165     & 1 \\
\safetysynth & (alternative) & 165    & 0 \\
\simpleBDD & (basic)        & 165     & 0 \\
\simpleBDD & (abs2) & 165     & 0 \\
\abssynthe & (SC3) 	 	    & 160     & 3 \\
\abssynthe & (SC2) 	 	    & 156     & 0 \\
\abssynthe & (SC1)					& 148			& 0 \\
\demiurge & (D1real)				& 127			& 11 \\
\termitesat	& 							& 101			& 6 \\
\bottomrule
\end{tabular}
}
\end{table}

\begin{figure}
\centering
\includegraphics[width=1\linewidth]{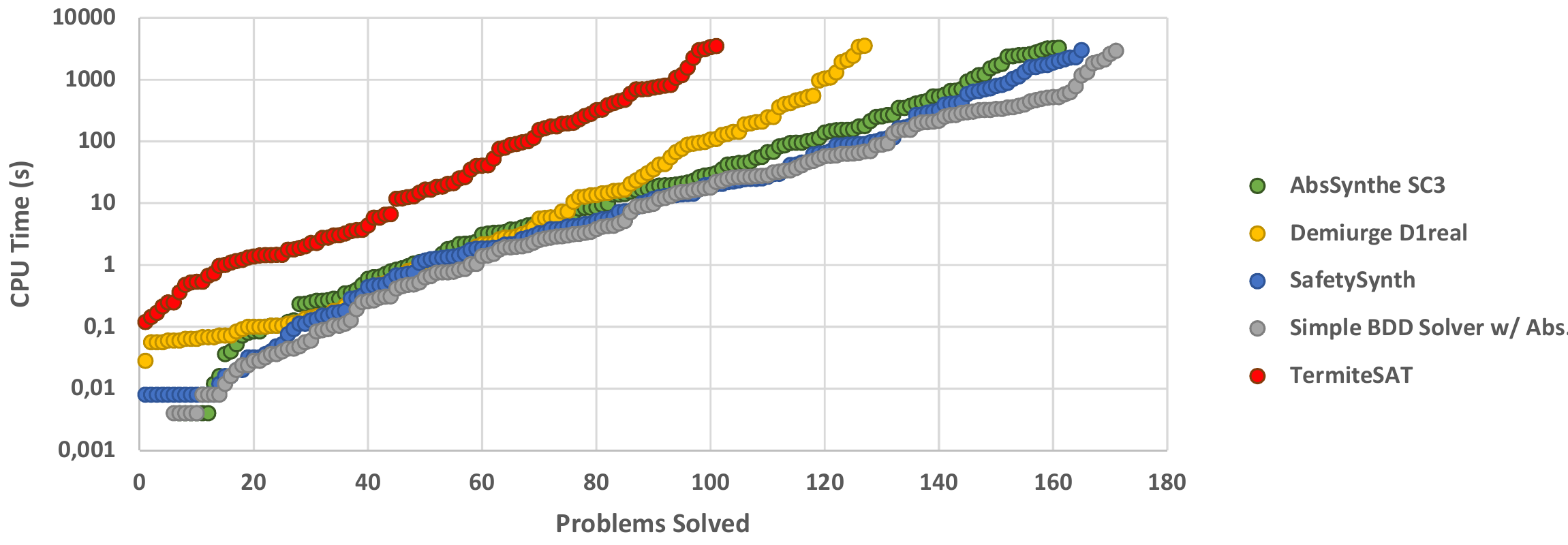}
\caption{Runtime Cactus Plot of Best Sequential Configurations}
\label{fig:cactus-realseq}
\end{figure}

Figure~\ref{fig:cactus-realseq} gives a cactus plot for the runtimes of the best sequential configuration of each tool.
%
%


\begin{figure}
\centering
\includegraphics[width=1\linewidth]{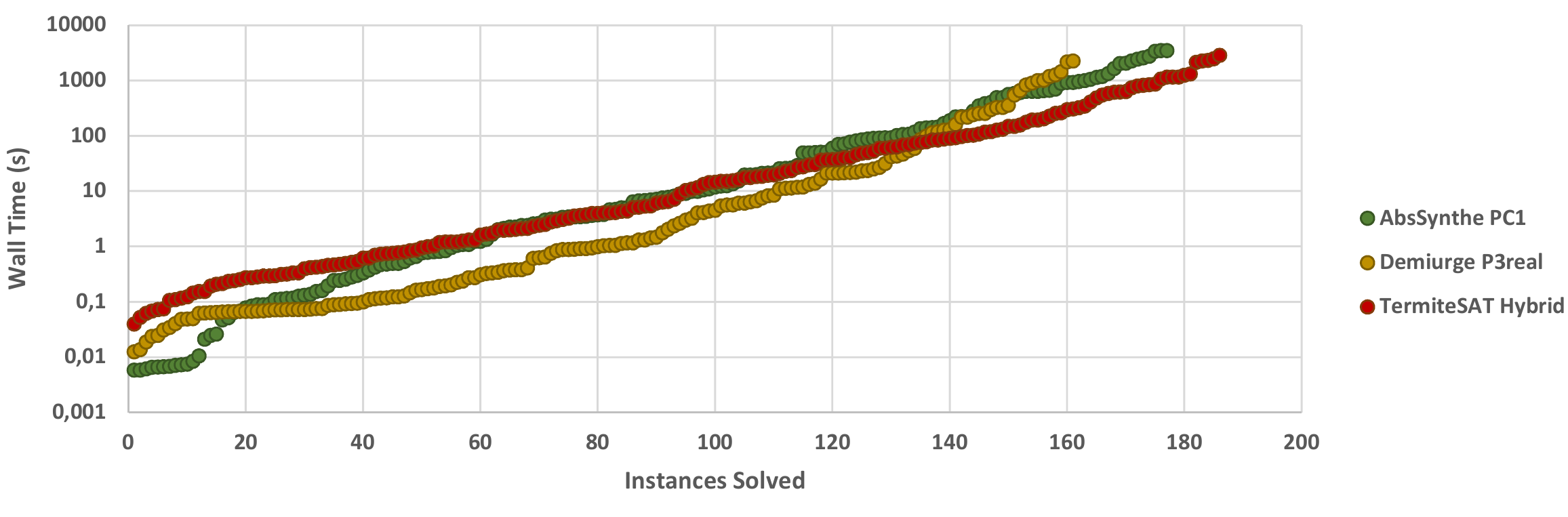}
\caption{Runtime Cactus Plot of Best Parallel Configurations}
\label{fig:cactus-realall}
\end{figure}	

\paragraph{Parallel Mode.}
Three of the tools that entered the competition had parallel configurations for the realizability track: three configurations of \abssynthe (par1, par2, par3), one configuration of \demiurge (P3real), and two configurations of \termitesat (portfolio, hybrid). These parallel configurations had to solve the same set of benchmark instances as in the sequential mode. In contrast to the sequential mode, runtime of tools is measured in wall time instead of CPU time. The results are given in Table~\ref{tab:results-realpar}. Compared to sequential mode, a number of additional instances could be solved: both \abssynthe and \termitesat have one or more configurations that solve more then the best tool in sequential mode (about $79\%$ of the benchmark set).
Only $15$ instances could not be solved by any tool in either sequential or parallel mode.

\begin{table}[h]
\caption{Results: AIGER Realizability (parallel mode only)}
\label{tab:results-realpar}
\centering
\def\arraystretch{1.3}
{\sffamily \small
\begin{tabular}{@{}llll@{}}
\toprule
Tool & (configuration)			& Solved  &  Unique \\ 
 \midrule
\termitesat & (hybrid) 		& 186 & 0 \\
\termitesat & (portfolio) & 185 & 0 \\
\abssynthe & (PC1) 	 	 	& 177 & 0 \\
\demiurge & (P3real) 			& 161 & 1 \\
\abssynthe & (PC3) 			& 156	& 0 \\
\abssynthe & (PC2)				& 147 & 0 \\
\bottomrule
\end{tabular}
}
\end{table}

Note that in Table~\ref{tab:results-realpar} we only count a benchmark instance as uniquely solved if it is not solved by any other configuration, including the sequential configurations. Consequently, only \demiurge (P3real) produces a single unique solution.

\begin{figure}[ht]
\centering
\includegraphics[width=\linewidth]{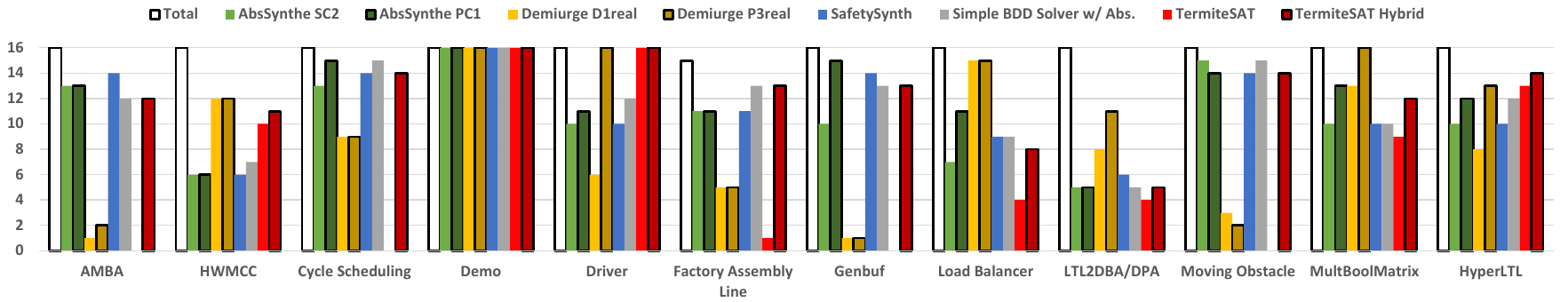}
%

\includegraphics[width=\linewidth]{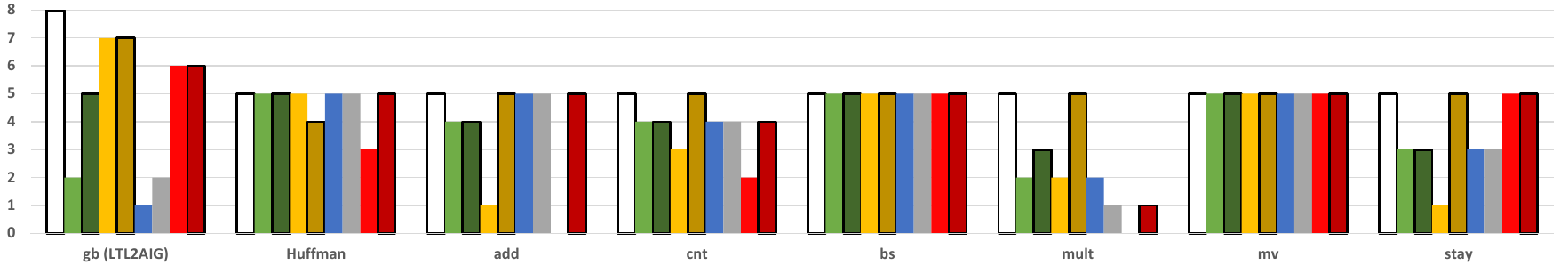}
\caption{AIGER/safety Realizability Track, Solved Instances by Category}
\label{fig:bycat2}
\end{figure}

\paragraph{Both modes: Solved Instances by Category.} Figure~\ref{fig:bycat2} 
gives an overview of the number of solved instances per configuration and category, for the best sequential and parallel configuration of each tool and the categories defined in Table~\ref{tab:selected-benchmarks}.

\paragraph{Analysis.}
The best sequential configuration on this year's benchmark set is \simpleBDD (abs1), the same as last year. This is the configuration with the simpler form of abstraction. Second place is shared between the other configurations of \simpleBDD and the two configurations of \safetysynth. Notably, \safetysynth (basic) is the only configuration that solves \texttt{amba16f110y}, one of the most difficult (unrealizable) benchmarks from the AMBA class.

These are followed by the three configurations of \abssynthe, where configuration (SC3), with compositionality and abstraction, performs best and solves 3 benchmarks uniquely (\texttt{genbuf64f100y}, \texttt{mult\_bool\_matrix\_6\_6\_8}, and \texttt{mult\_bool\_matrix\_6\_7\_7}), followed by configuration (SC2) which only uses abstraction, and then (SC1) which uses neither compositionality nor abstraction.
Like last year, \abssynthe still uses CUDD v2.5.1, compared to v3.0.0 used in \simpleBDD and \safetysynth. We observed last year that the switch to the newer version gave a significant speed-up to \simpleBDD. 

Finally, \demiurge and \termitesat again solve less problems than the BDD-based approaches, but solve a relatively large number of problems uniquely --- \demiurge mainly in classes HWMCC, Load Balancer, LTL2DBA/DPA, gb (LTL2AIG) and Huffman, and \termitesat mainly in HWMCC, HyperLTL and stay.

Among the parallel configuration, this year the two configurations of \termitesat solve most instances (with hybrid mode again only solving one additional instance compared to portfolio mode), followed by last year's winner \abssynthe (PC1) and the parallel configuration \demiurge (P3real). All of these show that a well-chosen portfolio can solve a significantly higher number of problems than the individual algorithms they are based on. 
Finally, the parallel configurations (PC2) and (PC3) of \abssynthe, which represent portfolios of (SC1) and (SC2), respectively, that only differ in the BDD reordering, solve roughly the same number as their sequential counterparts.

\subsection{AIGER/safety-Track: Synthesis}
In the track for synthesis from safety specifications in AIGER format, participants had to solve the same set of benchmarks as in the realizability track.
Three tools entered the track: \abssynthe, \demiurge and \safetysynth, in the same configurations as in the realizability track, except for additional synthesis of solutions.

For \syntcomp 2017, we have two different rankings in the synthesis track: one is based on the number of instances that can be solved within the timeout, and the other gives a weight to solutions of realizable specifications, based on their size. Furthermore, a solution for a realizable specification is only considered as correct if it can be model-checked within a separate timeout of one hour (cf. Section~\ref{sec:setup}).
As before, we first present the results for the sequential configurations, followed by parallel configurations, and end with an analysis of the results.

\paragraph{Sequential Mode.}
Table~\ref{tab:results-syntseq} summarizes the experimental results, including the number of solved benchmarks, the uniquely solved instances, the number of solutions that could not be model-checked within the timeout, and the accumulated quality of solutions. Note that the ``solved'' column gives the number of problems that have either been correctly determined unrealizable, or for which the tool has presented a solution that could be verified. With this requirement, no sequential configuration could solve more than $155$ or about $66\%$ of the benchmarks, and $48$ instances could not be solved by any tool.
Very few potential solutions could not be model-checked within the timeout. None of the tools produced any wrong solutions.\footnote{In the EDACC system, benchmark \texttt{amba16c40y} is reported with \texttt{model checking failed} for configuration \abssynthe (SC1). A closer inspection showed that this is not due to a faulty solution, but rather due to an uncaught exception in the model checker. Therefore, this solution is neither counted as an incorrect, nor as a correct solution.}


\begin{table}[h]
\caption{Results: AIGER Synthesis (sequential mode only)}
\label{tab:results-syntseq}
\centering
\def\arraystretch{1.3}
{\sffamily \small
\begin{tabular}{@{}llllll@{}}
\toprule
Tool & (configuration) & Solved & Unique & MC Timeout & Quality\\
\midrule
\safetysynth & (basic) 	& 155 & 2 & 0 & \textbf{236}\\
\safetysynth & (alt)		& 152 & 1 & 0 & 232\\
\abssynthe 	& (SC2) 		& 149 & 0 & 1 & 191\\
\abssynthe 	& (SC3) 		& 147 & 0 & 1 & 195\\
\abssynthe 	& (SC1) 		& 141 & 2 & 0 & 183\\
\demiurge 	& (D1synt) 	& 118 & 20 & 1 & 175\\
\bottomrule
\end{tabular}
}
\end{table}

\paragraph{Parallel Mode.}
Table~\ref{tab:results-syntpar} summarizes the experimental results, again including the number of solved benchmarks, the uniquely solved instances, the number of solutions that could not be verified within the timeout, and the accumulated quality of solutions. No tool solved more than $169$ problem instances, or about $72\%$ of the benchmark set. Again, there are only very few (potential) solutions that could not be verified within the timeout. None of the solutions have been identified as wrong by our verifiers.

Like in the parallel realizability track, we only consider instances as uniquely solved if they are not solved by any other configuration, including sequential ones. Consequently, none of the configurations produced any unique solutions.


\begin{table}[h]
\caption{Results: AIGER Synthesis (parallel mode only)}
\label{tab:results-syntpar}
\centering
\def\arraystretch{1.3}
{\sffamily \small
\begin{tabular}{@{}llllll@{}}
\toprule
Tool & (configuration) & Solved & Unique & MC Timeout & Quality\\
\midrule
\abssynthe  & (PC1) 		& 169 & 0 & 2 & 210\\
\demiurge 	& (P3Synt) 	& 158 & 0 & 0 & \textbf{266}\\
\abssynthe 	& (PC3) 		& 148 & 0 & 2 & 198\\
\abssynthe 	& (PC2) 		& 139 & 0 & 1 & 179\\
\bottomrule
\end{tabular}
}
\end{table}
\paragraph{Analysis.}
Unsurprisingly, the number of solved instances for each tool in the synthesis track corresponds roughly to those solved in the realizability track. With respect to the smaller number of participants, \demiurge (D1synt) provides an additional $20$ unique solutions in the sequential mode, and a few unique solutions are also provided by \abssynthe (SC1) and \safetysynth (basic and alt). 

\begin{figure}[h]
\centering
\includegraphics[width=\linewidth]{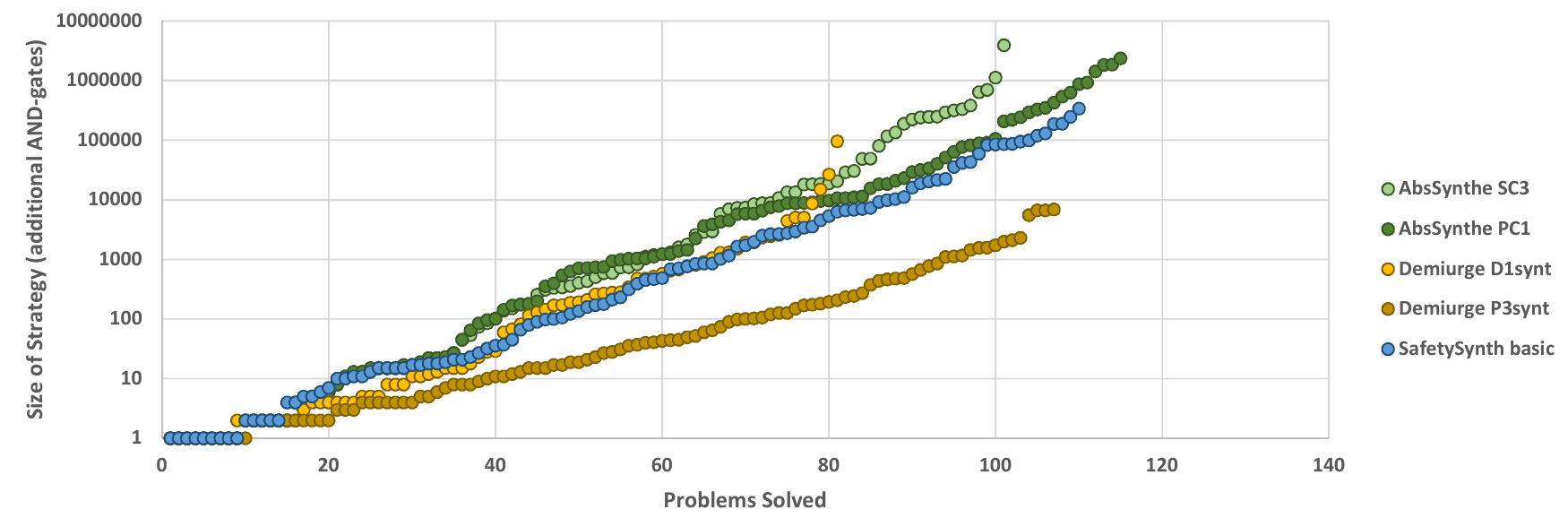}
\caption{AIGER/safety Synthesis Track: Size of Solution Strategies for Selected Configurations}
\label{fig:size-cactus}
\end{figure}

In sequential mode, the tool with the overall highest quality of solutions is the one which provides also the highest number of solutions, \safetysynth (basic). Note also that here \abssynthe (SC2) solves more problems than (SC3), but (SC3) has a higher quality score, which means that on average it gives better solutions than (SC2). In parallel mode, the quality score of \demiurge (P3synt) is about $25\%$ higher than for \abssynthe (PC1), even though the latter solves about $7\%$ more problems. This means that on average the solutions of \demiurge are significantly smaller than those produced by \abssynthe and \safetysynth.
This can be seen in Figure~\ref{fig:size-cactus}, which  plots the sizes of synthesized strategies for some of the configurations. We consider here not the size of the complete solution (which includes the specification circuit), but only the number of additional AND-gates, which corresponds to the strategy of the controller.

%% file: results-tlsf.tex
\subsection{TLSF/LTL-Track: Realizability}

In the track for realizability checking of LTL specifications in TLSF, $5$ tools competed in $8$ sequential and $5$ parallel configurations. In the following, we compare the results of these $13$ configurations on the $244$ benchmarks that were selected for \syntcomp 2017, as explained in Section~\ref{sec:selection}. 

Again, we first restrict our evaluation to sequential configurations, then extend it to include parallel configurations, and finally give a brief analysis.

\paragraph{Sequential Mode.}
In sequential mode, \acaciaforaiger, \bowser and \ltlsynt each competed with one configuration\footnote{In fact, \bowser was entered also in the realizability track in three configurations, but these configurations differ only in the synthesis step. Therefore, they all produced identical results, and are represented as a single configuration here.}, \bosy with two configurations (ltl3ba and spot), and \party with three configurations.

The number of solved instances per configuration, as well as the number of uniquely solved instances, are given in Table~\ref{tab:results-realseq-tlsf}. No tool could solve more than $218$ out of the $244$ instances, or about $89\%$ of the benchmark set\footnote{For two configurations of \party, the numbers reported here differ from those given by the EDACC system. This is because configurations (bool) and (aiger) give result \texttt{unrealizable} on benchmarks \texttt{ltl2dba\_R\_10} and \texttt{ltl2dba\_R\_12} only after an uncaught error that is highlighted by the solver output. Therefore, these two (correct) results are not counted here.} $14$ instances could not be solved by any of the participants within the timeout. 

\begin{table}[h]
\caption{Results: TLSF Realizability (sequential mode only)}
\label{tab:results-realseq-tlsf}
\centering
\def\arraystretch{1.3}
{\sffamily \small
\begin{tabular}{@{}llll@{}}
\toprule
Tool & (configuration)		& Solved  &  Unique \\ 
\midrule
\party 					& (aiger) & 218 & 7 \\
\ltlsynt				& 						& 195 & 3 \\
\bosy 					& (spot) 			& 181	& 0 \\
\bosy						& (ltl3ba) 		& 172	& 0 \\
\party					&	(int)				& 169	& 0 \\
\bowser					& 						& 165	& 0 \\
\party					& (bool)			& 164	& 0 \\
\acaciaforaiger	& 						& 142	&	4 \\
\bottomrule
\end{tabular}
}
\end{table}

\begin{figure}[h]
\centering
\includegraphics[width=1\linewidth]{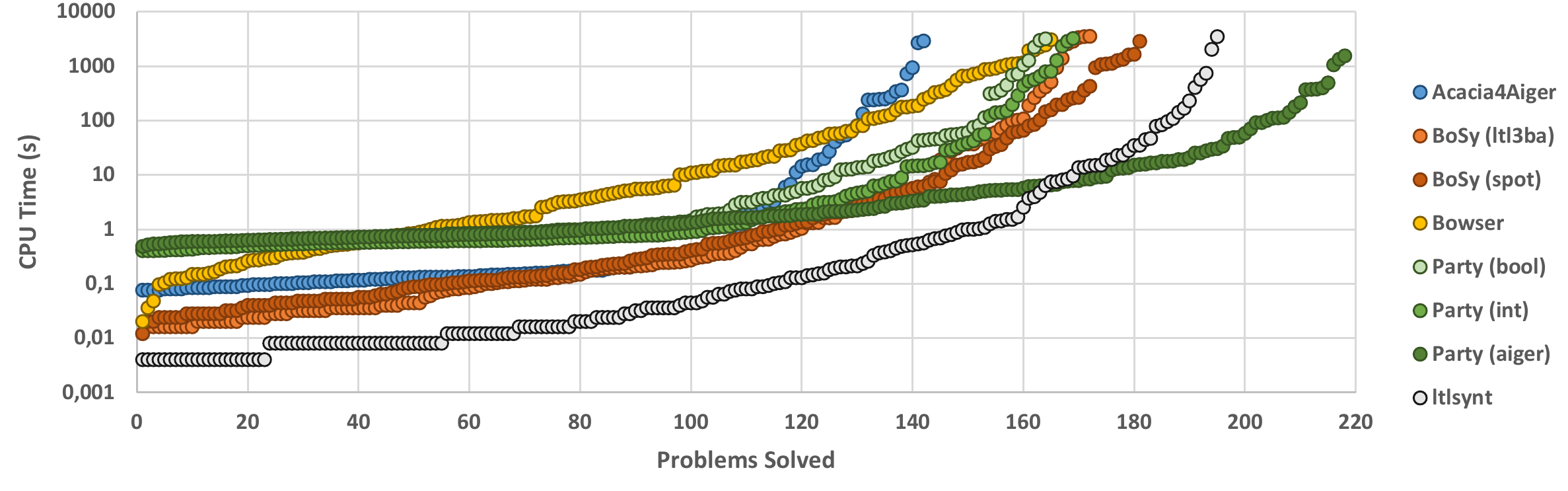}
\caption{TLSF/LTL Realizability Track: Runtimes of Sequential Configurations}
\label{fig:cactus-realseq-tlsf}
\end{figure}

Figure~\ref{fig:cactus-realseq-tlsf} gives a cactus plot of the runtimes for all sequential algorithms in the realizability track.
%
%


\paragraph{Parallel Mode.}
All tools except \ltlsynt also entered in one or more parallel configurations: one configuration for each of \acaciaforaiger, \bowser and party, and two configurations for \bosy. As before, parallel configurations solve the same set of benchmark instances as in the sequential mode, but runtime is measured in wall time instead of CPU time. The results are given in Table~\ref{tab:results-realpar-tlsf}. The best parallel configuration solves $224$ out of the $244$ instances, or about $92\%$ of the benchmark set. $10$ benchmarks have not been solved by any configuration.

\begin{table}
\caption{Results: TLSF Realizability (parallel mode only)}
\label{tab:results-realpar-tlsf}
\centering
\def\arraystretch{1.3}
{\sffamily \small
\begin{tabular}{@{}llll@{}}
\toprule
Tool & (configuration)			& Solved  &  Unique \\ 
 \midrule
\party 		& (portfolio)		& 224	& 2\\
\bosy 		& (spot,par)				& 181	& 0\\
\bowser		& (par)							& 173	& 0\\
\bosy			& (ltl3ba,par)			& 170	& 0\\
\acaciaforaiger & (par)				& 153	& 0\\
\bottomrule
\end{tabular}
}
\end{table}

In Table~\ref{tab:results-realpar-tlsf}, we again only count a benchmark instance as uniquely solved if it is not solved by any other sequential or parallel configuration. Then, \party (portfolio) solves $2$ instances that no other configuration can solve.

\begin{figure}[h]
\centering
\includegraphics[width=1\linewidth]{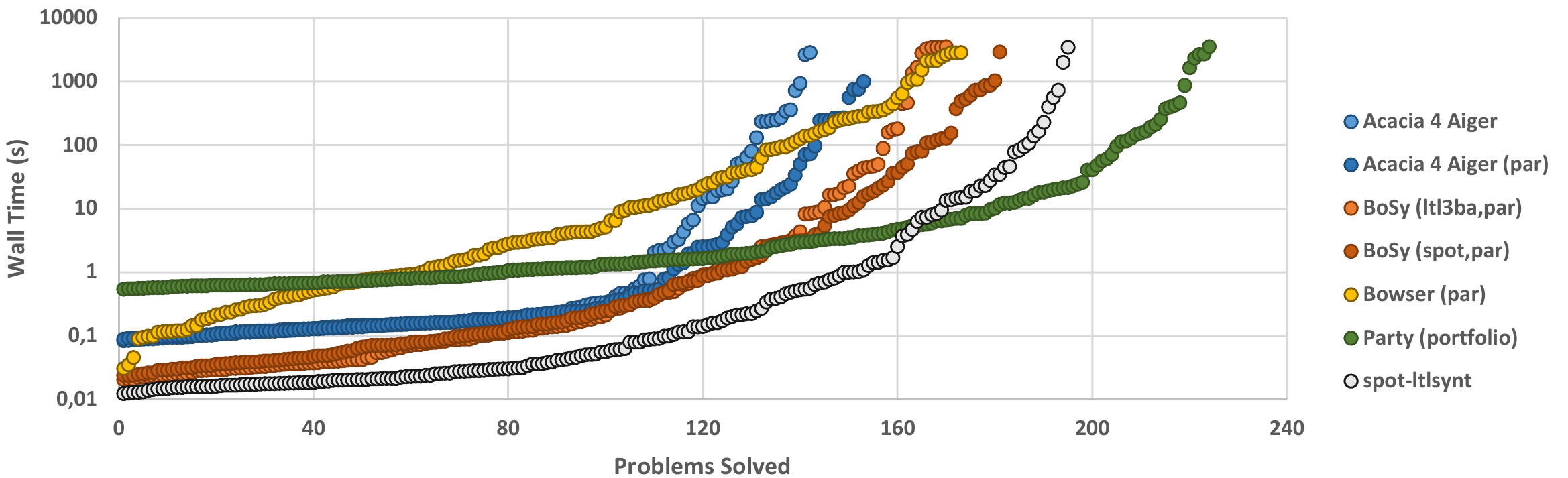}
\caption{TLSF/LTL Realizability Track: Runtimes of Parallel and Selected Sequential Configurations}
\label{fig:cactus-realall-tlsf}
\end{figure}

Figure~\ref{fig:cactus-realall-tlsf} gives a cactus plot of the runtimes for the parallel and a selection of the sequential algorithms in the realizability track.

\begin{figure}[h]
\centering
\includegraphics[width=\linewidth]{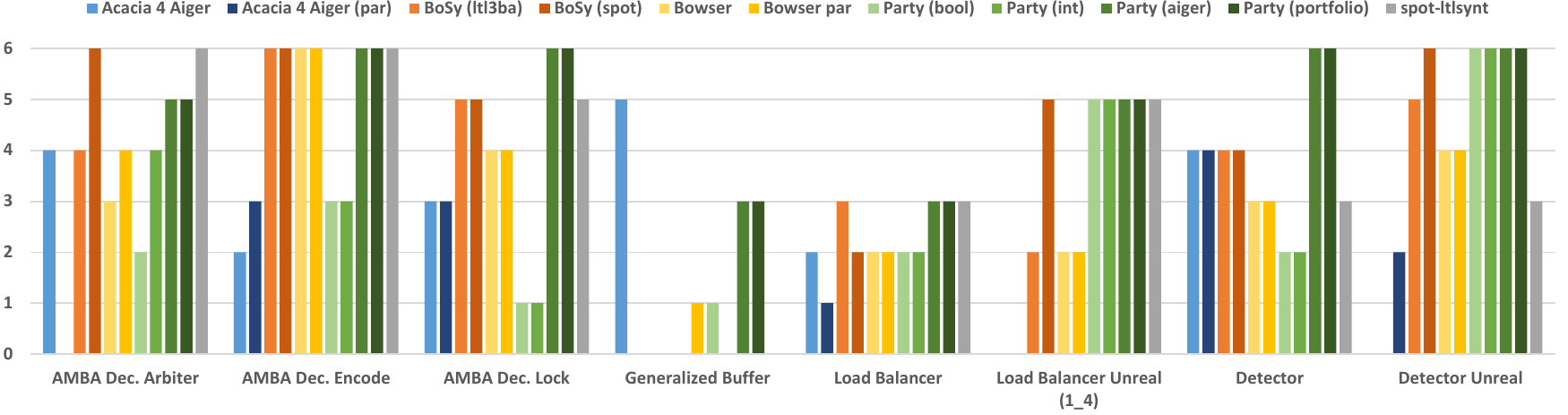}
%
\vspace{1em}

\includegraphics[width=\linewidth]{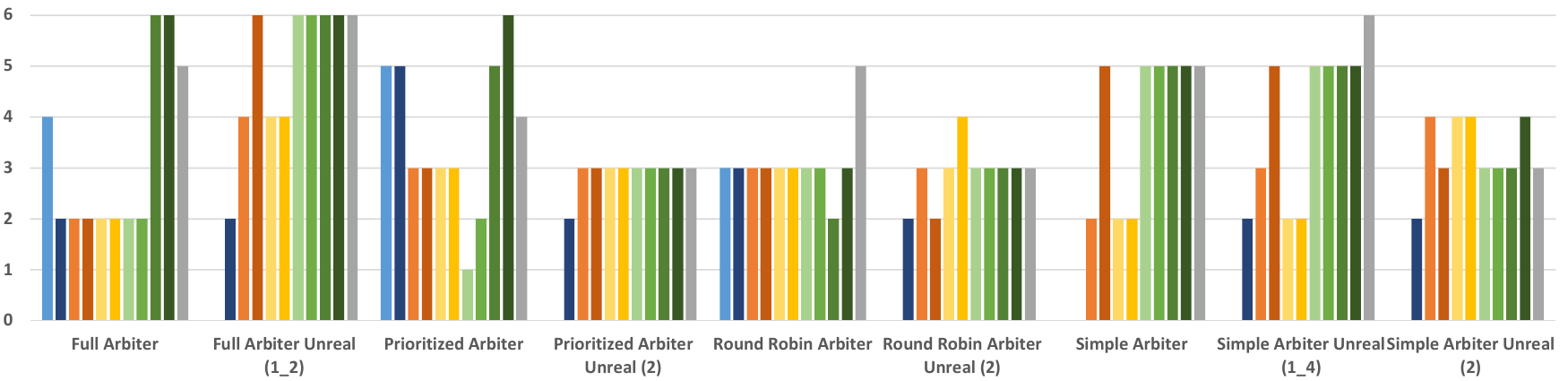}
\vspace{1em}

\includegraphics[width=\linewidth]{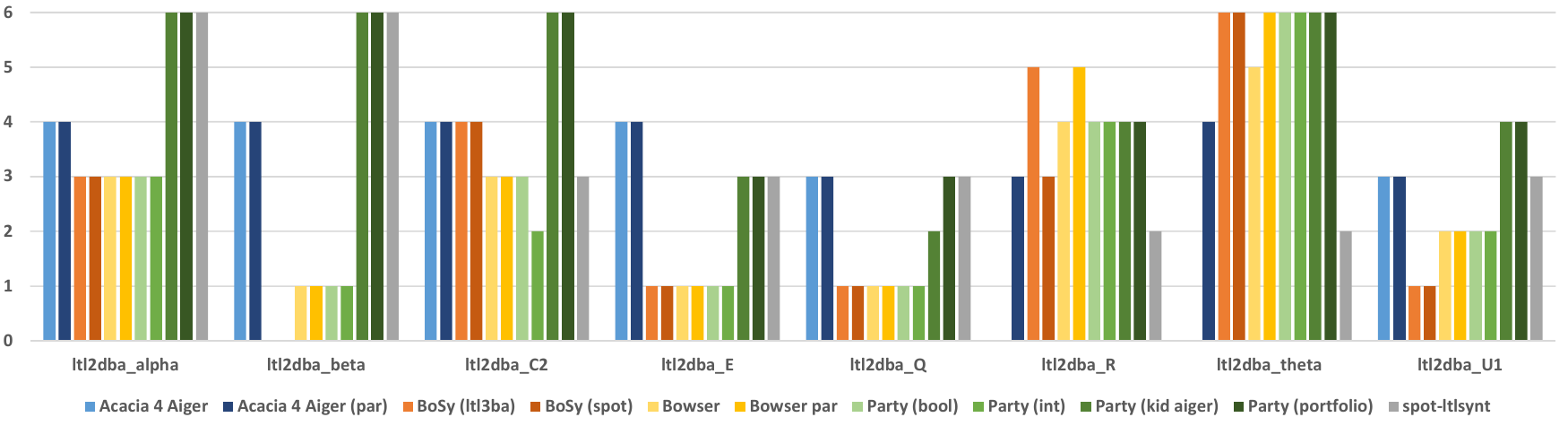}
\caption{TLSF/LTL Realizability Track: Solved Instances for Parameterized Benchmarks}
\label{fig:bycat-tlsf}
\end{figure}

\paragraph{Both modes: Solved Instances of Parameterized Benchmarks.} 
For both the sequential and the parallel configurations, Figure~\ref{fig:bycat-tlsf} gives an overview of the number of solved instances per configuration, for the $25$ parameterized benchmarks used in \syntcomp 2017.

\paragraph{Analysis.}
In contrast to last year, where the competitors essentially only implemented two different synthesis approaches, the entrants of \syntcomp 2017 are much more diverse: \bosy, \bowser and \party all implement variants of bounded synthesis by encoding into a constraint satisfaction problem, and four other approaches are implemented in \acaciaforaiger, \ltlsynt, and \party (bool and aiger).\footnote{Note that \acaciaforaiger had some technical problems with the syntax of a subset of the new benchmarks for this year, so the number of solved instances does not necessarily reflect the number of instances that could be solved with the implemented approach in principle.} Remarkable are the two completely new approaches implemented in \party (aiger) and \ltlsynt, which both solve more problems than the best tools from last year. \party (aiger) achieves this by a translation to safety games which approximates liveness properties by bounded $k$-liveness. In contrast to all other approaches, \ltlsynt does not work by a conversion to a bounded liveness or bounded size problem, which allows the other approaches to avoid determinization of specification automata. Instead, \ltlsynt achieves very good results with an algorithm based on deterministic automata and parity games, which was widely assumed to be impractical before. We conjecture that this relies on at least two factors: i) very efficient translation and determinization algorithms implemented in Spot~\cite{Duret16,Redziejowski12}, and ii) the choice of automata with transition-based acceptance, because their conciseness is key to the efficiency of the overall algorithm.

An analysis of the parameterized benchmarks in Figure~\ref{fig:bycat-tlsf} shows the different strengths of the approaches: \acaciaforaiger dominates on \texttt{generalized\_buffer} and \texttt{ltl2dba\_E}, the bounded synthesis-based approaches on \texttt{ltl2dba\_R} and \texttt{simple\_arbiter\_unreal2}, the (aiger) configuration of \party on \texttt{amba\_decompsed\_lock}, \texttt{detector}, \texttt{full\_arbiter}, \texttt{ltl2dba\_C2} and \texttt{ltl2dba\_U1}, and \ltlsynt on \texttt{round\_robin\_arbiter} and \texttt{simple\_arbiter\_unreal1\_4}. For the other parameterized benchmarks, several approaches share the top spot, with the (aiger) configuration of \party almost always among the best.

For the benchmarks that have been made unrealizable by adding additional requirements, we also note that the different differences between the approaches: compared to the realizable version \texttt{simple\_} \texttt{arbiter}, the first unrealizable version \texttt{simple\_arbiter\_unreal1\_4} is comparably hard for most tools and easier for some. In contrast, the second unrealizable version \texttt{simple\_arbiter\_unreal2} has a different behavior for most tools, being either significantly harder or significantly easier.

Finally, we note that the two instances that are solved uniquely by \party (portfolio) (\texttt{prioritized\_} \texttt{arbiter\_7.tlsf} and \texttt{simple\_arbiter\_12.tlsf}) are solved by the second portfolio solver with formula strengthening, which did not compete in the sequential mode.

\subsection{TLSF/LTL-Track: Synthesis}
In the track for synthesis from LTL specifications in TLSF, participants had to solve the same benchmarks as in the LTL/TLSF-realizability track.
Up to the \bowser tool, the track had the same participants as the LTL/TLSF-realizability track: \acaciaforaiger with two configurations, \bosy with four configurations, \party with four configurations, and \ltlsynt with one configuration. \bowser competed in six configurations.

As for the AIGER/safety-track, there are two rankings in the synthesis subtrack, one based on the number of instances that can be solved, and the other based on the quality of solutions, measured by their size. Again, a solution for a realizable specification is only considered correct if it can be model-checked within a separate timeout of one hour (cf. Section~\ref{sec:setup}).
We start by presenting the results for the sequential configurations, followed by parallel configurations, and end with an analysis of the results.

\paragraph{Sequential Mode.}
Table~\ref{tab:results-syntseq-tlsf} summarizes the experimental results for the sequential configurations, including the number of solved benchmarks, the uniquely solved instances, and the number of solutions that could not be model-checked within the timeout. The last column gives the accumulated quality points over all correct solutions.

As before, the ``solved'' column gives the number of problems that have either been correctly determined unrealizable, or for which the tool has presented a solution that could be verified. With this requirement, no sequential configuration could solve more than $200$ or about $82\%$ of the benchmarks, and $29$ instances could not be solved by any tool. None of the tools provided any wrong solutions.

In this track, we note that all configurations that are not based on some form of bounded synthesis generated a significant number of solutions that could not be verified.


\begin{table}[h]
\caption{Results: TLSF Synthesis (sequential mode only)}
\label{tab:results-syntseq-tlsf}
\centering
\def\arraystretch{1.3}
{\sffamily \small
\begin{tabular}{@{}llllll@{}}
\toprule
Tool & (configuration) & Solved & Unique & MC Timeout & Quality\\
\midrule
\party 					& (aiger) & 200 & 4 & 20 & 219 \\
\ltlsynt				& 						& 182 & 3 & 13 & 180 \\
\bosy 					& (spot) 			& 181	& 3 & 0  & \textbf{298}\\
\party					&	(int)				& 167	& 0 & 0  & 249\\
\bosy						& (ltl3ba) 		& 165	& 0 & 0  & 277\\
\party					& (bool)			& 163	& 1 & 0  & 222\\
\bowser					& (c0)				& 162	& 0 & 0  & 273\\
\bowser					& (c1)				& 155	& 0 & 0  & 260\\
\acaciaforaiger	& 						& 110	&	2 & 17 & 91\\
\bowser					& (c2)				& 93	& 0 & 0  & 141\\
\bottomrule
\end{tabular}
}
\end{table}

\paragraph{Parallel Mode.}
In this mode, \acaciaforaiger, \bosy and \bowser competed with parallel version of their configurations from the sequential track. Additionally, \party competed in a portfolio approach that combines its sequential configurations.
Table~\ref{tab:results-syntpar-tlsf} summarizes the experimental results, in the same format as before. No configuration solved more than $203$ problem instances, or about $83\%$ of the benchmark set. $27$ benchmarks could not be solved by any tool. \party and \acaciaforaiger produced a significant number of solutions that could not be verified within the timeout. None of the solutions were determined to be wrong.

As before, we only consider instances as uniquely solved if they are not solved by any other configuration, including sequential ones. Consequently, none of the solutions are unique. 

\begin{table}[h]
\caption{Results: TLSF Synthesis (parallel mode only)}
\label{tab:results-syntpar-tlsf}
\centering
\def\arraystretch{1.3}
{\sffamily \small
\begin{tabular}{@{}llllll@{}}
\toprule
Tool & (configuration) & Solved & Unique & MC Timeout & Quality\\
\midrule
\party 					& (portfolio) & 203 & 0 & 18 & \textbf{308} \\
\bosy 					& (spot,par) 	& 181	& 0 & 0  & 297\\
\bosy						& (ltl3ba,par)& 169	& 0 & 0  & 286\\
\bowser					& (c0,par)		& 169	& 0 & 0  & 285\\
\bowser					& (c1,par)		& 169	& 0 & 0  & 285\\
\bowser					& (c2,par)		& 168	& 0 & 0  & 290\\
\acaciaforaiger	& (par)				& 137	&	0 & 5  & 123\\
\bottomrule
\end{tabular}
}
\end{table}

\begin{figure}[!h]
\centering
\includegraphics[width=\linewidth]{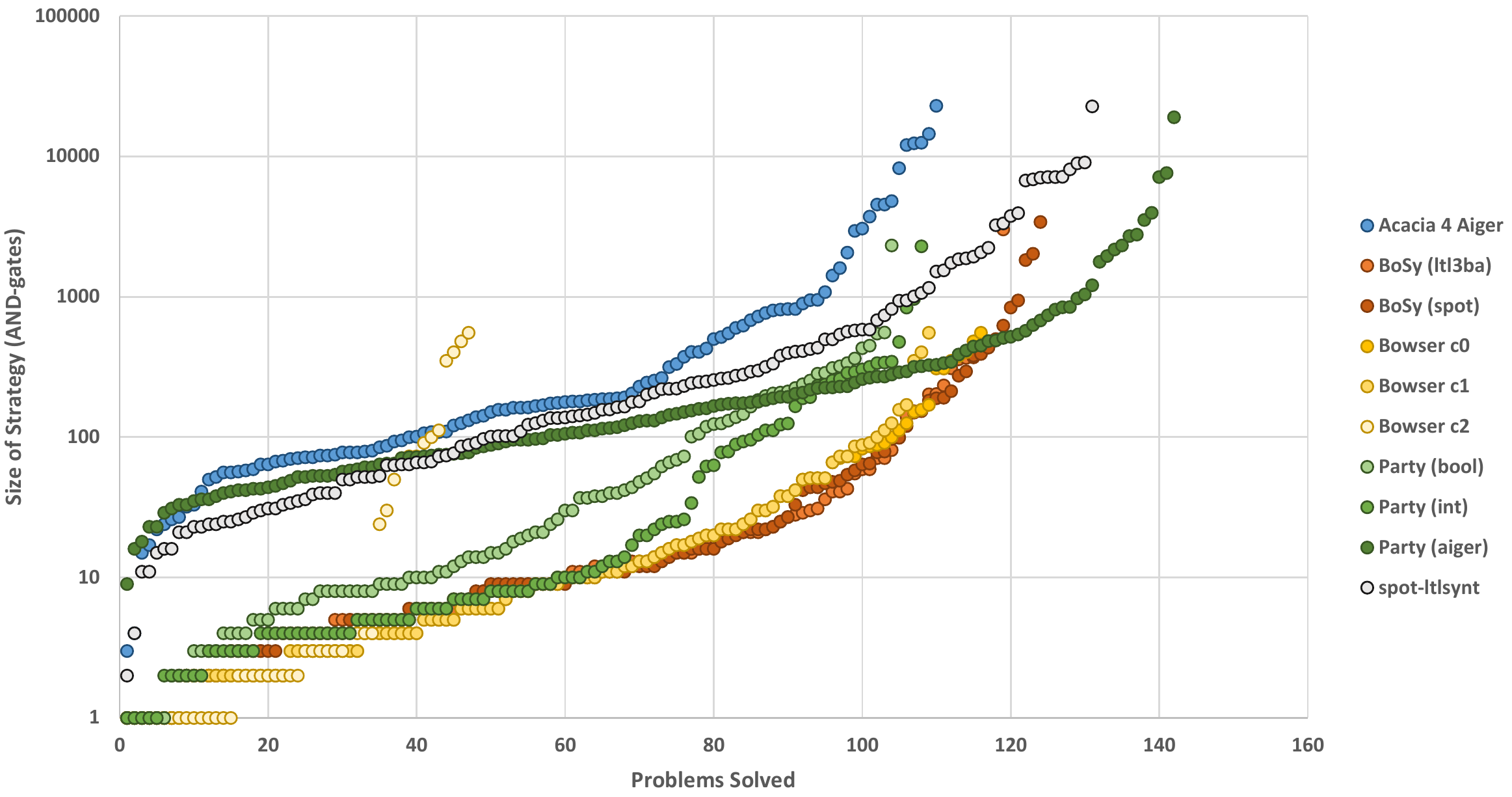}
\caption{TLSF/LTL Synthesis Track: Solution Sizes of Sequential Configurations}
\label{fig:TLSF-size1}
%
\centering
\includegraphics[width=\linewidth]{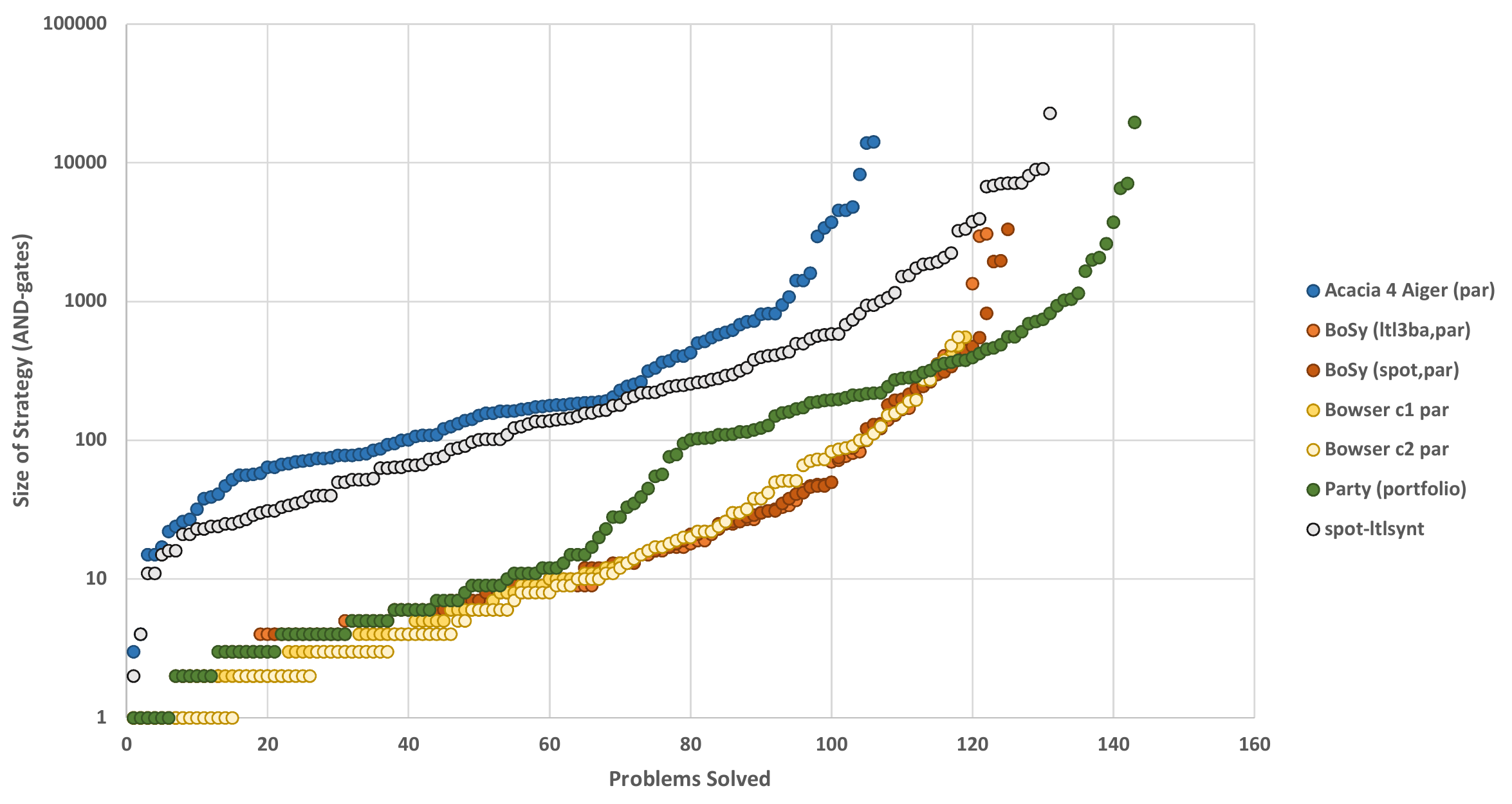}
\caption{TLSF/LTL Synthesis Track: Solution Sizes of Parallel Configurations and \ltlsynt}
\label{fig:TLSF-size2}
\end{figure}

\paragraph{Analysis.}
As for the AIGER/safety-track, the number of solved instances for each tool in synthesis is closely related to the solved instances in realizability checking. The number of unique instances of \acaciaforaiger and \party (aiger) decreases, in part due to solutions that could not be verified. As a consequence, we also note that \bosy (spot) and \party (bool) now have some unique solutions, i.e., they do not provide the only solution, but the only solution that could be verified. 

Considering the quality ranking, \bosy (spot) is the best tool in the sequential subtrack: even though it produces about $10\%$ fewer solutions than \party (aiger), its accumulated quality points are about $36\%$ higher. In fact, in the ranking based on quality, all bounded synthesis-based configurations are better than \party (aiger), except for \bowser (c2). 
A different picture unfolds in the parallel subtrack, were \party (aiger) combines the strengths of its different approaches to find high-quality solutions for at least some of the benchmarks, such that it not only solves the highest number of problems, but also achieves the highest accumulated quality.

Figure~\ref{fig:TLSF-size1} plots the sizes of sequential configurations. It shows, as expected, that the bounded synthesis approaches produce much smaller solutions than the other approaches. In particular, the solution sizes of all configurations of \bosy and \bowser are very similar, with \bowser producing more very small solutions (with $<10$ AND-gates), and \bosy producing slightly better solutions for the remaining problems.  The approach of \party (bool) falls somewhere in between. It also shows that the approach of \party (aiger) in many cases produces significantly smaller solutions than the approaches of \acaciaforaiger and \ltlsynt. 

In parallel mode, depicted in Figure~\ref{fig:TLSF-size1}, we can see changes mostly for \bowser (c2,par) and \party (portfolio). The latter manages to combine the strengths of its different approaches, providing small solutions for those problems that can still be solved by its (int) configuration, and otherwise falling back to the (aiger) configuration. \bowser (c2,par) shows the strength of its approach to produce the smallest possible solutions: for more than $60$ benchmarks it provides the smallest solution. 

A further analysis on the quality and the size of implementations shows that \bowser (c2,par) is the configuration that has the highest average quality for the problems that it does solve. Furthermore, it produces the highest number of new reference solutions, i.e., solutions that have a quality greater than $2$ according to the quality ranking scheme explained in Section~\ref{sec:rules}. The analysis for all tools is given in Table~\ref{tab:quality-synt-tlsf}.

\begin{table}[h]
\caption{TLSF Synthesis: Average Quality and New Reference Solutions}
\label{tab:quality-synt-tlsf}
\centering
\def\arraystretch{1.3}
{\sffamily \small
\begin{tabular}{@{}llll@{}}
\toprule
Tool & (configuration) & Avg. Quality & New Ref. Solutions \\
\midrule
\bowser					& (c2,par)		& 1.725 & 50\\
\bosy						& (ltl3ba,par)& 1.691 & 31\\
\bowser					& (c1,par)		& 1.689 & 40\\
\bowser					& (c0,par)		& 1.688 & 40\\
\bowser					& (c0)				& 1.686 & 37\\
\bosy						& (ltl3ba)		& 1.679 & 20\\
\bowser					& (c1)				& 1.676 & 34\\
\bosy 					& (spot) 			& 1.644 & 30\\
\bosy 					& (spot,par) 	& 1.643 & 23\\
\party 					& (portfolio) & 1.517 & 27\\
\bowser					& (c2)				& 1.514 & 20\\
\party 					& (int) 			& 1.493 & 20\\
\party 					& (bool) 			& 1.363 & 15\\
\party 					& (aiger) & 1.093 & 13\\
\ltlsynt				& 						& 0.988	& 8\\
\acaciaforaiger	& (par)				& 0.898 & 0\\
\acaciaforaiger	& 						& 0.825 & 0\\
\bottomrule
\end{tabular}
}
\end{table}

%% file: conclusions.tex
\section{Conclusions}
\label{sec:conclusions}

\syntcomp 2017 consolidated the changes made last year, most importantly the introduction of the track for LTL specifications in the temporal logic synthesis format (TLSF). This year, two completely new tools have been entered in the competition: \bowser and \ltlsynt. Furthermore, four tools have received (sometimes major) updates, and four tools have been re-entered in the same version as last year. The only major change to the rules is the re-introduction of a quality ranking for the synthesis tracks.

In the AIGER/safety tracks, we had rather small changes on the tools and on the benchmark set, and this is reflected in similar, but not identical results as last year: \simpleBDD (abs1) again wins the sequential realizability mode, and the parallel realizability mode this year goes to \termitesat (hybrid), which was a close second to \abssynthe last year. In the synthesis track, \safetysynth (basic) and \abssynthe (PC1) again solve most problems in the sequential and parallel mode, respectively. In the quality ranking, \safetysynth (basic) also is the best configuration in sequential mode, and \demiurge (P3synt) is the best in parallel mode.

In the TLSF/LTL tracks, we had significant changes to both the tools and the benchmark set, including two new tools and a large number of new benchmarks. Consequently, the results look a lot different than last year. In fact, all of the winners are tools or configurations that did not participate last year: in sequential realizability and sequential synthesis, the new configuration (aiger) of \party solves most problems, followed by the new tool \ltlsynt. In parallel realizability and synthesis, the new configuration \party (portfolio) solves most problems. Finally, in the quality ranking, new configuration \bosy (spot) has the highest accumulated quality in sequential mode, and \party (portfolio) wins the parallel mode.

{ \small
\myparagraph{Acknowledgments}
The organization of \syntcomp 2016 was supported by the Austrian Science Fund
(FWF) through project RiSE (S11406-N23) and by the German
Research Foundation (DFG) through project ``Automatic Synthesis of 
Distributed and
Parameterized Systems'' (JA 2357/2-1), and its setup and execution by the 
European Research Council (ERC) Grant OSARES (No.~683300). 

The development of \abssynthe and \acaciaforaiger was supported by an F.R.S.-FNRS and FWA fellowships, and
the ERC inVEST (279499) project.
%

The development of \safetysynth and \bosy was supported by the 
ERC Grant OSARES (No.~683300). 
%
%
}